\documentclass[journal=jpcbfk,manuscript=article, longbibliography]{achemso}

\usepackage{bm}%
\usepackage{graphicx}
\usepackage{float}
\usepackage{setspace}%
\usepackage{fullpage}
\graphicspath{ {Figures/} }
\usepackage{color,soul}
\usepackage[normalem]{ulem}
\usepackage[usenames,dvipsnames]{xcolor}

\title{Active Learning the Coarse-Grained Energy Landscape For Water Clusters From Sparse Data}
\author{Troy D. Loeffler}
\author{Tarak K. Patra}
\author{Henry Chan}
\author{Mathew Cherukara}
\author{Subramanian K.R.S. Sankaranarayanan}
 
\affiliation{Department of Mechanical and Industrial Engineering, University of Illinois,Chicago, Illinois 60607, United States
\newline
Center for Nanoscale Materials, Argonne National Lab, Lemont, Illinois 60439, United States}
\email{tloeffler@anl.gov, skrssank@anl.gov, skrssank@uic.edu}
\date{April 2019}

\begin{document}

\maketitle

\begin{abstract}
Molecular dynamics with pre-defined functional forms is a popular technique for understanding dynamical evolution of systems. The pre-defined functional forms impose limits on the physics that can be captured. Artificial Neural Network (ANN) models have emerged as an attractive flexible alternative to the expensive quantum calculations (e.g. density functional theory) calculations in the area of molecular force-fields. Ideally if one is able to train a ANN to accurately predict the correct DFT energy and forces for any given structure, they gain the ability to perform molecular dynamics with high accuracy while simultaneously reducing the computation cost in a dramatic fashion. While this goal is very lucrative, neural networks are interpolative and therefore it is not always clear how one should go about training a neural network to exhaustively fit the entire phase space of a given system. Currently, ANNs are trained by generating large quantities (On the order of $10^{4}$ or greater) of structural data in hopes that the ANN has adequately sampled the energy landscape both near and far-from-equilibrium. This can, however, be a bit prohibitive when it comes to more accurate levels of quantum theory. As such it is desirable to train a model using the absolute minimal data set possible, especially when costs of high-fidelity calculations such as CCSD and QMC are high. Here, we present an Active Learning approach that iteratively trains an ANN model to faithfully replicate the coarse-grained energy surface of water clusters using only 426 total structures in its training data. Our active learning workflow starts with a sparse training dataset which is continually updated via a Monte Carlo scheme that sparsely queries the energy landscape and tests the network performance. Next, the network is retrained with an updated training set that includes failed configurations/energies from previous iteration until convergence is attained. Once trained, we generate an extensive test set of ~100,000 configurations sampled across clusters ranging from 1 to 200 molecules and demonstrate that the trained network adequately reproduces the energies (within mean absolute error (MAE) of ~ 2 meV/molecule) and forces (MAE ~ 40 meV/{\r{A}}) compared to the reference model. More importantly, the trained ANN model also accurately captures both the structure as well as the free energy as a function of the various cluster sizes. Overall, this study reports a new active learning scheme with promising strategy to develop accurate force-fields for molecular simulations using extremely sparse training data sets.

\end{abstract}

\section{Introduction}

Classical molecular dynamics (MD) is a powerful technique for materials modeling and is typically used to understand the dynamical evolution of nanoscale systems. The accuracy of MD hinges strongly on the accuracy of the potential model used to describe the interactions between the atoms in the system. Recently, MD has shown its predictive prowess in several different applications from tribology to catalysis. Its ability to capture the underlying physics and chemistry, however, is limited by the pre-defined functional form of the potential model. A greater flexibility in the functional form of MD is highly desirable.

Artificial Neural Networks (ANN) have emerged as a viable flexible alternative to the pre-defined empirical functional forms.  ANNs are a set of mathematical equations whose structure is designed to be a software mimic of the synaptical network found within the brains of living organisms. While the concept of an ANN has been around for many decades\cite{Specht1991}, it was not up until recently that their wide spread usage became feasible due to massive improvements in computational hardware. ANN have become an attractive option to predict the density functional theory (DFT) energies and forces of an arbitrary structure without the computational overhead that normal DFT calculations entail. Due to it's functional complexity, the same neural network is able to take on a wide variety of curvatures depending on the choice of weights in the network. This can allow it to fit high dimensional potential energy surfaces with a fair degree of accuracy (typically within chemical accuracy ~ 1-10 meV/atom).\cite{Behler2014} This has been used to fit a wide variety of systems such as Zinc Oxide\cite{Nongnuch2011}, Copper\cite{Artrith2012}, Germanium Teleride\cite{Sosso2012}, and other systems.\cite{Handley2014, ARTRITH2015} However, neural networks are interpolative and the high degree of flexibility can also mean that a poor choice of weights or network structure can result in some incredibly wild trends and can even result in nearly discontinuous behavior. As a result, properly training of a neural network and avoiding overfitting is a challenging task. Often, even if a researcher is able to reliably train the network for one set of material or conditions, the transferability of the network to other material types and/or conditions is often questionable since neural networks typically produce poor predictions when extrapolating outside of their training data. As such, how a network is trained is of utmost importance.

Given their interpolative nature, training a neural network to predict configurational energies/forces in molecular systems often requires large quantities of training data, typically on the order of $10^{4}$ to $10^{5}$ structures or greater. Such an extensive training data is used to ensure that the the model has enough of a basis to reasonably train a model.\cite{Boes2016} For reasonably "cheap" levels of quantum theory such as DFT or MP2 generating this many structures is still in the realm of feasibility. However, as one goes to more expensive calculations such as Coupled Cluster or Quantum Monte Carlo, generating a training set of this size begins to become computationally intractable. In addition, even when one provides this many structures, it is still possible to leave out or under represent important structures that can play an important role in accurately predicting the thermophysical properties of a given system. Ideally, one would like to be able to fit a neural network with minimal apriori knowledge of all of the relevant structures. 

In order to address this challenge, we introduce an Active Learning (AL) scheme that samples the potential energy landscape  on-the-fly and without human intervention makes a decision on the structures that need to be included in the training set. Most importantly, this AL scheme starts with minimalist number of training data (i.e. 1 point) and iteratively builds up the smallest training set that best represents the energy landscape. Using a representative coarse-grained system of water clusters, we demonstrate that the AL scheme can train a high quality neural network potential to correctly replicate the energy surface and many of the physical properties of a complicated 3 body reference potential while at the same time using minimal amount of training data.

\section{Method}

A typical AL iteration contains the following major steps. 
\begin{enumerate}
  \item Training of the NN using the current structure pool.
  \item Running a series of stochastic algorithms to test the trained network's current predictions 
  \item An identification of configurational space where the NN is currently struggling 
  \item An update of the structure pool with failed configurations
  \item Retraining of the NN with the updated pool and back to step 2
\end{enumerate}

The workflow described above is schematically depicted in Figure 1. To test our AL scheme, we train a neural network to a reference Tersoff\cite{Tersoff1989,Tersoff1988,munetoh2007interatomic} based \cite{chan2019machine} coarse grained water model, which contains a two and three body term within it's potential. The neural networks used in this study were constructed and trained using the Atomic Energy Network (AENet) software package\cite{AENet2016}, which was modified to implement the active learning scheme outlined above. Simulations using these networks were carried out using AENet interfaces with the Classy Monte Carlo simulation software to perform the AL iterations. 

For the structure of the network, a set of Belher\cite{Behler2011} style symmetry function were used to construct the invariant coordinates. We use 8 radial symmetry functions, $G^{2}$ and 18 angular symmetry functions, $G^{5}$ for constructing the coordinate space.  The parameters used for each function can be found in Table 1. Two layers of 10 twist style nodes were used for the hidden layers. A Levenberg-Marquardt approach was used to optimize the neural network weights for each AL generation. This was done with a batch size of 32 structures and a learn rate of 0.1 once the structure pool was large enough to accommodate these settings. Initially, the batch size was set to 1, given the small initial training data set. For each network generation, the neural network is trained for a total of 2,000 epochs, where each epoch represents one complete training cycle. AENet makes use of a k-fold cross validation scheme, where a given fraction (k) of the training set is not used for the objective minimization. Instead this fraction is used to cross validate the training process to minimize over-fitting. For each AL iteration, the network which had the the best error from the cross validation was chosen as the best network for this AL iteration and is carried forward. 

Once the best network has been chosen, a series of simulations are run to actively sample the configurational space predicted by the current neural network. It was found that MD is not suitable for sampling within this scheme due to the fact that when the network is still in its infancy, large spikes in the forces can lead to unphysical acceleration of particles within the simulation box. In addition, even in a reasonably well trained network, MD can be trapped in a local energy well that prevents it from searching the phase space outside of this well. This can often create models that work well within the trained local minima, but can have catastrophically bad predictions when the model is applied to environments found outside of the training set. Monte Carlo and other similar sampling methods in contrast are much less sensitive to spikes in the energy surface which make them more suitable methods for sampling poorly trained energy landscapes. 

In addition, a wide collection of non-physical moves or non-thermal sampling approaches can be used. For the purposes of this work, Boltzmann based Metropolis sampling and a nested ensemble based approach\cite{Nielsen2013} were used to generate the structures for each AL iteration. This was done to gather information on both thermally relevant structures predicted by the neural network as well as higher energy structures which may still be important for creating an accurate model. The Metropolis simulation was ran for 5,000 MC cycles at 300K with the initial structure being randomly picked from the current neural network training pool. The Nested Ensemble simulations were ran for another 5,000 cycles.

After the stochastic sampling step is completed, a set of 10 structures are gathered from the trajectory files of the Metropolis and Nested Sampling files. The real energy of these structures are computed and compared. For each structure, if the neural network prediction and the exact energy do not agree within a given tolerance, the structure is then added to the training pool to be used for the next AL iteration. 
This entire process is continued until the exit criteria is hit. For this work, we specified that if no new structures were added in 5 consecutive AL iterations, that the potential has converged. For the addition tolerance, we specified that any structure with a greater than difference of 1 meV between the real and predicted energy should be added to the training pool.

Since an initial neural network can not be trained on zero data, a single structure is used to seed the initial neural network in order to kick off the training process. This was chosen to be a reasonably minimized structure in order to ensure at least one low energy configuration was contained in the training set. Theoretically one could begin with any number of seed structures, but for the purposes of evaluating the efficiency of this approach, the absolute minimal seed data was used. In order to rigorously validate the neural network models, we created a test set that consists of roughly 140,000 cluster structures ranging in size from 2 molecules all the way up to 70.

\section{Results}
We first evaluate the performance of our active learning (AL) scheme. Figure 2 (a) shows the mean absolute error (MAE) in meV/atom as a function of the AL iterations; each iteration corresponds to an epoch or complete training cycle. The corresponding number of structures added for each AL iteration is also shown in Fig. 2 (a). As mentioned earlier, the initial NN is trained with minimal number of training configurations. Hence, it is not surprising that the MPE has high errors ~ 115 meV/atom. These errors drop progressively with AL iterations i.e. from ~115 meV/atom at iteration 1 to less than 5 meV/atom at iteration 140, as more distinct (failed) structures are added to the pool. The initial errors, of course, were nearly on the same magnitude as the total system energy of ~200-400 meV/atom. As the AL algorithm progressed, the error temporarily plateaued out ~ 40 meV/atom, which may suggest that the NN search hit a local minima. Eventually the error sees a sharp drop till it reaches the stopping criteria with the final network having a total mean error of 1.2 meV/molecule. After about a total of 145 total AL iterations, the system had finally reached the stopping criteria i.e. no new structures are added during five consecutive test cycles. At the time when this criteria was hit, the final structure count had reached a total of 426 unique training structures. 

To measure how well the networks performed as a function of the number of AL iterations, the best network from each Al iteration was taken and used to predict the energy on a test set that comprises of 140000 configurations and their energies. A correlation plot of the network vs reference energy data for the final optimized network on the training data set generated over the course of the active learning iterations is shown in Fig.\ref{NetVsReal_Energy}. As expected, we find that the final optimized network is able to reliably predict the cluster energies for the training data set generated not only near equilibrium, but also in the highly non-equilibrium region that extends far beyond (i.e. -0.2 meV/atom and higher). It is worth noting that the MAE of the NN on the training data is ~0.38 meV/atom, which is well within chemical accuracy.

We generate a test set of 140,000 clusters to evaluate the performance of the optimized network. The MAE on the total test comprising of all the clusters was ~ 2 meV/atom. We next compare the performance of the NN over a range of cluster sizes. A correlation plot showing the predictions of the NN energies as a function of cluster size with respect to the target is shown in Figs. 3 (a)-(c). We observe that the NN performs very well across the entire range of clusters sampled; the MAE were found to be 2.58, 2.18, and 1.15 meV/atom for the 1-20, 21-50, and 51+ ranges respectively. While the errors are generally small, it appears from the observed trend that the error declined as the cluster size grew larger and larger as shown in Fig. 3d. The highest errors were found for cluster sizes between 2 and 7 molecules, which had an average error of ~ 2.5 meV/molecule. As the size of the cluster increased beyond 20 molecules in size, the MAE gradually declined down to a value of around 1.0 meV/atom. The slightly higher errors for the smaller sized clusters is understandable considering that surface and finite-size effects are more prominent.

In addition to checking the energy predictions of the neural network, we also evaluate the performance of the NN on forces. Note that the forces were not included as part of the training during the AL iterations. Correlation plots comparing the which can be found in Figs. \ref{NetVsReal_Force} (a)-(c). The overall MAE between the reference and neural network was found to be 41.45 meV/\r{A}. Considering the network had not been trained on the forces, the agreement was found to be of excellent quality. We further evaluate the performance for force predictions as function of cluster size; correlation plots for the 1-20, 21-50, and 51+
are given in (a), (b) and (c), respectively.  Each point in the correlation plot is given as the difference between the Fx, Fy, and Fz components for every atom in the structure. The MAE was found to be 27.48, 40.14, and 42.30 meV/A for the 1-20, 21-50, and 51+ ranges. Overall the errors received from this network for the test set are quite satisfactory over a wide range of clusters beyond what the network was trained against. 

We note that simply comparing single snapshot energies is not good enough to deduce that a model is good or bad since a model may often fall apart when predictions of the physical properties are made with it. As such it is prudent to begin studying different thermo-physical properties and evaluate the NN performance against more rigourous thermophysical properties such as free energies. For a cluster system, especially, one of the most difficult properties to correctly predict is the free energy as a function of cluster size. This property is dependent not only on correctly predicting the minimal energy configuration of a cluster, but also in correctly predicting the high energy yet thermally relevant states. We compute the free energy using the EB-AVBMC method\cite{Loeffler2015}. The results from these calculations are shown in Fig. \ref{ClusterFreeEnergy}. Overall, we find that there is a reasonable agreement between the neural network prediction and the original Tersoff model although the neural network tends to slightly underpredict the free energy for large size clusters. These values are very difficult to get exactly due to the fact that the free energy for a size of N is also dependent on the free energy of all the sizes smaller than N. Although the free energy of the clusters are not explicitly trained to the network, the AL-ANN is able to produce moderate agreement between the reference potential and the AL-ANN prediction for water clusters of varying size. 

We next compare the structural property using our ANN and the reference model is shown in Fig. \ref{ClusterRDF}. Specifically, we compute the radial distribution function (RDF) for clusters containing 10, 30, and 50 water molecules. The results are shown in Fig. \ref{ClusterRDF} (a)-(c), respectively. The agreement between the peak/trough locations and heights were all found to be within a 1\% error which implies the RDF was faithfully replicated by the AL-ANN model. Overall, given the sparse training data set that comprised of only configurations vs. energies, we find that the AL-ANN was able to reasonably reproduce many of the structural and physical properties of original reference model.

It is worth noting that one can make an argument that for this particular energy surface, it is likely any 500 structures may be sufficient to train an ANN to correctly replicate the data shown here. To illustrate that this is not the case, we randomly generate a set of 500 structures consisting of a cluster size of 25 (similar to the size sampled in the AL pool) using the reference Tersoff potential and using sampling based on the Metropolis MC and Nesting Ensemble techniques used in the work flow. A network using the same network structure as for the AL, was trained using these 500 structures and compared against the same test data set of 140000 clusters. The results of these are shown in Fig. S1 in the Supporting Information. From these results, we find that the actual number of structures is not as important as the number of unique structures contained within the training pool. In addition, generating structures from the exact potential using directed sampling such as a Boltzmann sampling, appears to not be sufficient. This will  often leave out higher energy structures that, while not being thermally relevant, are nonetheless important for teaching the ANN, the correct trends and behaviors.

Finally, we note that though this training procedure provided excellent results, it is still possible to extend it even further. The fixed cluster size of 25 molecules used in the training process for this study can be relaxed to include a wider variety of cluster sizes. This can become more important when attempting to fit DFT and ab-initio data where polarizability and delocalization effects become more dominant. Furthermore, we note that, in the present work, the structure of the network remained fixed for a single AL run. However, this generally assumes that one knows the best choice of symmetry functions that they will need for a given system. Using the wrong choice of network structure or symmetry functions can produce poor results even with a good training procedure.\cite{Nongnuch2011} To circumvent this, one can perform AL iterations on the network hyperparameters as well to obtain the best possible network while simultaneously reducing the size of the training data employed.

\section{Conclusion}

In conclusion, we have introduced an automated active learning workflow for training ANN. We demonstrate that our AL scheme allows for on-the-fly sampling of the configurational and potential energy surface and produces a high-quality neural network with a sparse dataset that comprises of only 426 structures. We further show that our network was able to make excellent predictions of both the energies and forces over a wide variety of cluster sizes (test set ~ 140000 configurations) that were not originally part of the training set. In addition, the actively learned network model was able to predict thermophysical properties such as free energy as a function of cluster size as well as correctly replicate the radial distribution functions for clusters containing 10, 30, and 50 molecules. Considering that high-fidelity quantum calculations such as quantum Monte Carlo (QMC) and coupled clusters (CCSD) are computationally expensive, one can only generate sparse data sets. In this context, we believe that our AL scheme overcomes one of the major limitations of training NN against sparse datasets and lays the groundwork for future training of ANN against sparse high-fidelity data from quantum calculations.

\section{Acknowledgements}
Use of the Center for Nanoscale Materials and the resources of the Argonne Leadership Computing Facility was also supported by the U. S. Department of Energy (DOE), Office of Science, Office of Basic Science, under the contract no. DE-AC02-06CH11357. This research used resources of the National Energy Research Scientific Computing Center, a DOE office of science user facility supported by the Office of Science of the US Department of Energy under contract no.  DE-AC02-05CH11231. This work was supported by Argonne LDRD-2017-012-N0.

\bibliography{Refs}

\newpage
\begin{table}[h]
\includegraphics[scale=0.6]{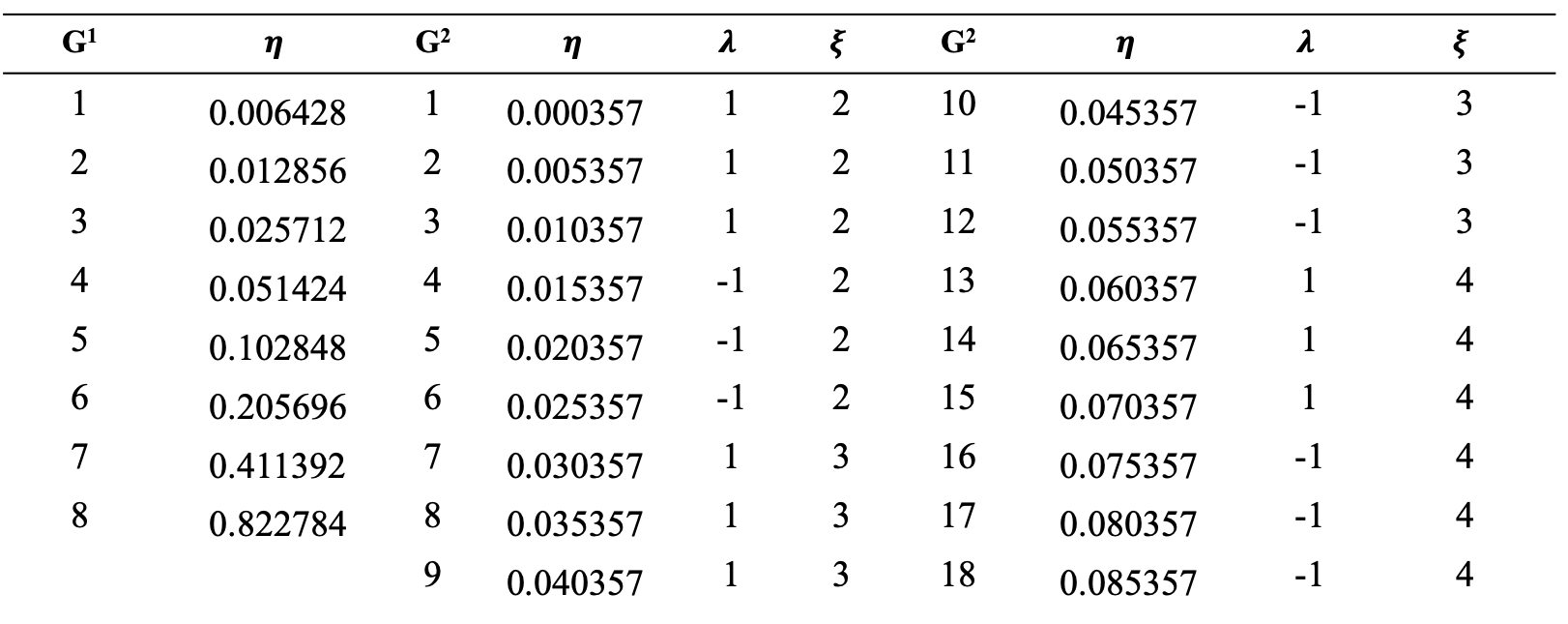}
\caption{\label{G2G4} Parameters of the 8 radial symmetry functions $G^{2}$ and 18 angular symmetry functions $G^{5}$ with a cut-off distance of 3.5$\AA$}
\end{table}
\pagebreak[4]

\newpage
\begin{figure}[h]
\includegraphics[scale=0.6]{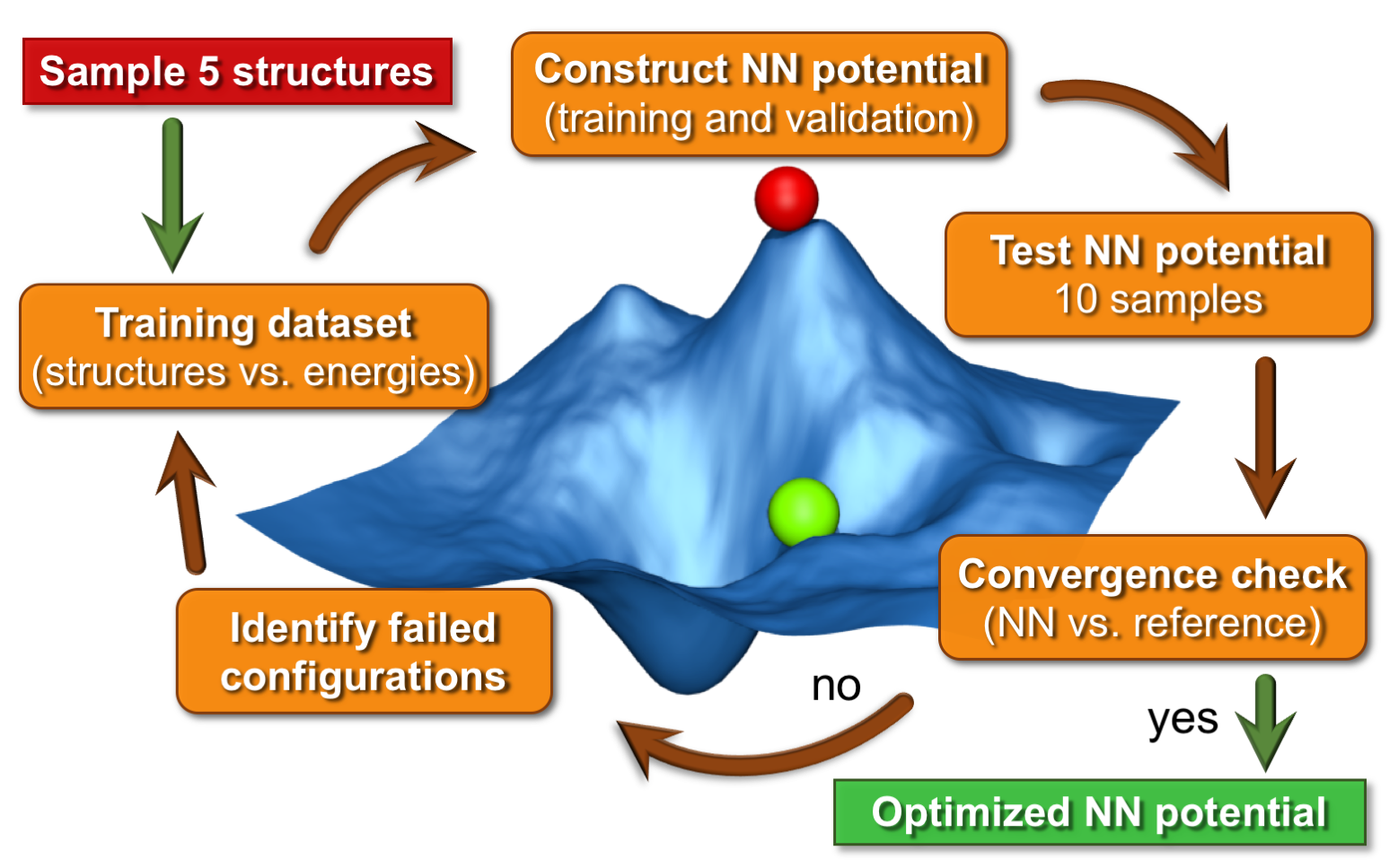}
\caption{\label{ALFlowChart} The workflow of the Active Learning scheme}
\end{figure}
\pagebreak[4]

\newpage
\begin{figure}[h]
\includegraphics[trim={2cm 5cm 2cm 2cm},scale=0.57]{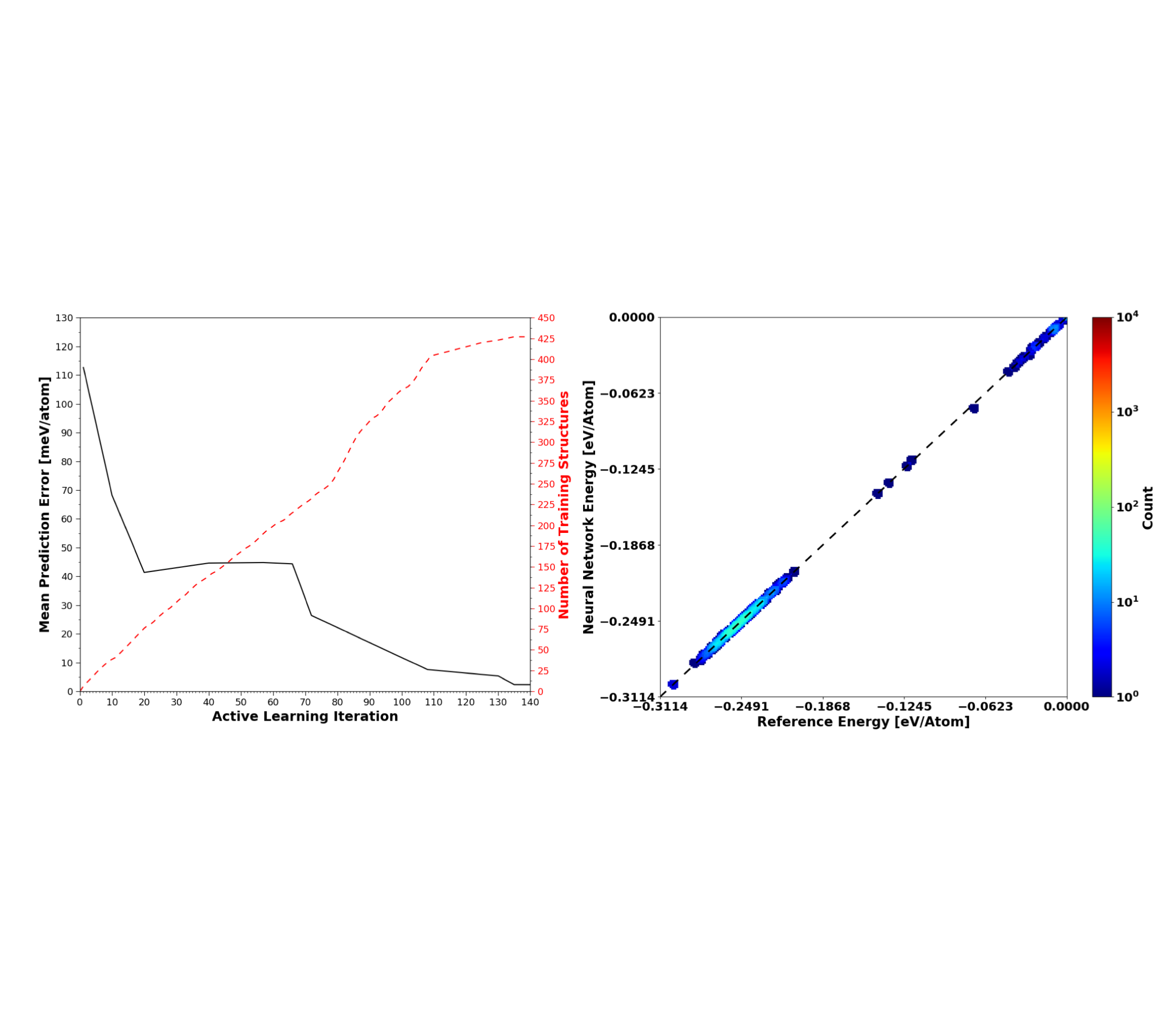}
\caption{\label{ErrorVsGen} Active learning of the NN potential for water. (a) the mean absolute error of the AL-ANN tested on the 140,000 cluster validation set is given as a function of active learning iteration or generation (solid black curve). RHS of the same plot shows the size of the training data (dashed red curve) for the same training generation. (b) A correlation plot showing the performance of the final optimized network on the 426 structure training set. The NN predictions of energy are compared against the target energy. The mean absolute error for the training set was found to be 0.38 meV/atom.}
\end{figure}
\pagebreak[4]

\newpage
\begin{figure}[h]
\includegraphics[trim={4cm 1cm 1cm 1cm},scale=0.56]{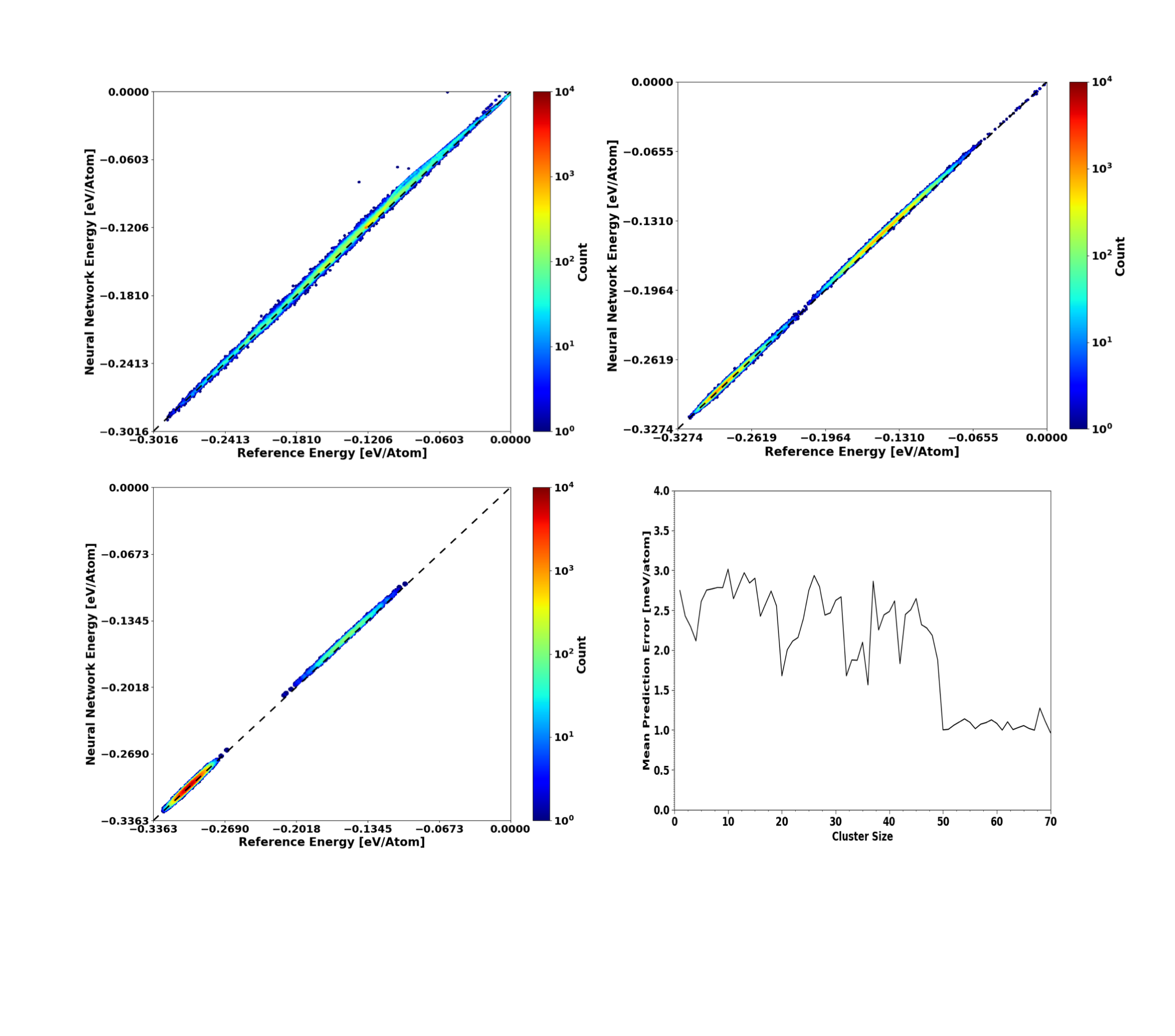}
\caption{\label{NetVsReal_Energy}  Energy correlation plots showing the evaluation of the network performance as a function of cluster size. (a) Clusters containing 1-20 water molecules (b) clusters containing 20-50 water molecules and  (c) clusters containing 51+ water molecules.  For the overall test set comprising of all the cluster, the total mean absolute error was  found to be 2.01 meV/atom. The MAE were found to be 2.58, 2.18, and 1.15 meV/atom for the 1-20, 21-50, and 51+ ranges, respectively. The MAE in predicted energy is shown as a function of cluster size in (d).}
\end{figure}
\pagebreak[4]

\newpage
\begin{figure}[h]
\includegraphics[trim={4cm 1cm 1cm 1cm},scale=0.65]{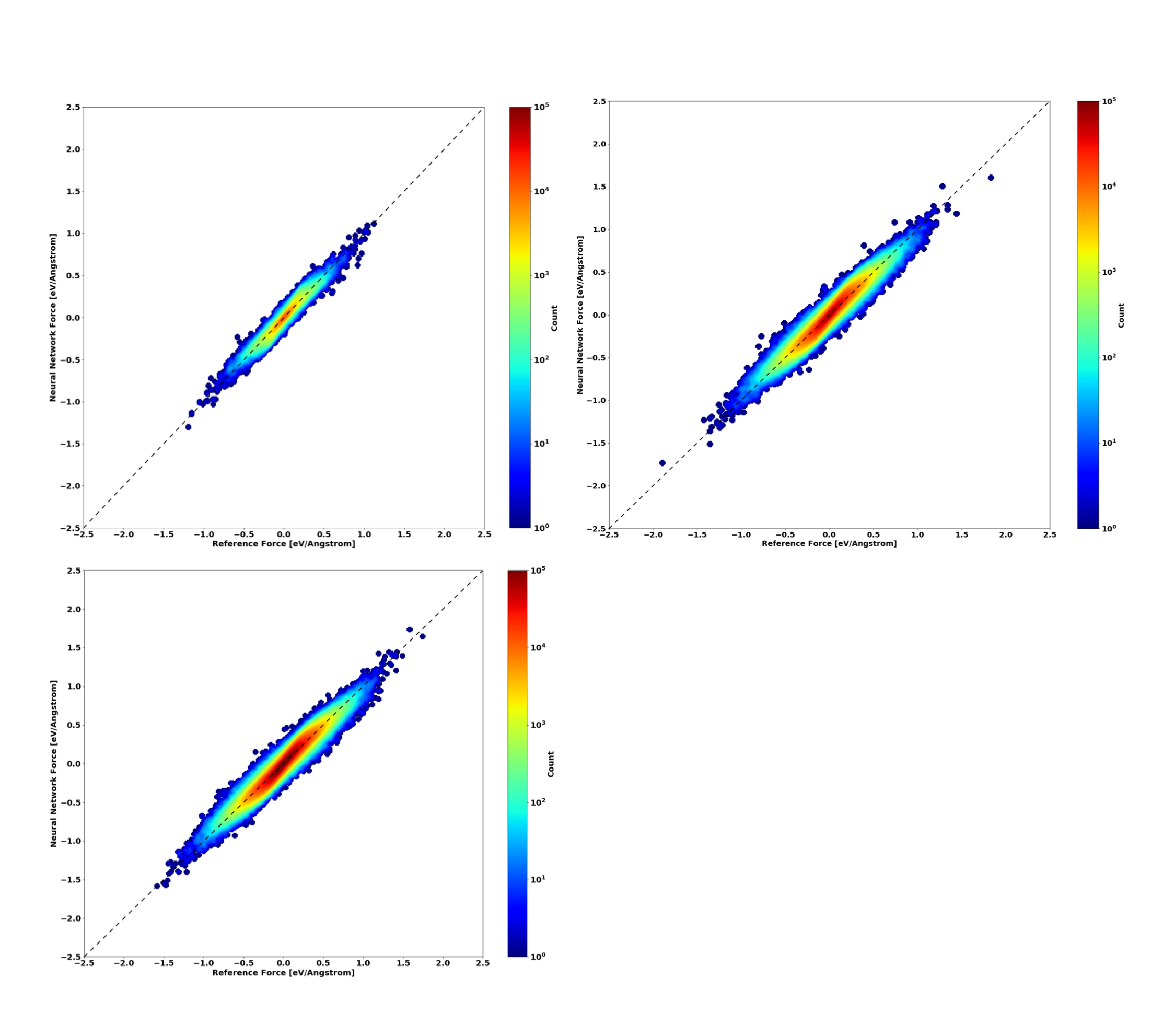}
\caption{\label{NetVsReal_Force} Force correlation plots showing the evaluation of the network performance in predicting forces as a function of the cluster size. The final network's predicted force for test set comprising of 140,000 structures is compared against that predicted by the reference model. The results for the 1-20, 21-50, and 51+ are given in (a), (b) and (c), respectively. Each point in the correlation plot is the difference between the $F_{x}$, $F_{y}$, and $F_{z}$ components for every atom in the structure. The mean absolute error for was found to be 27.48, 40.14, and 42.30 meV/\r{A} for the 1-20, 21-50, and 51+ ranges, respectively. For the entire set the error was found to be 41.45 meV/\r{A}}
\end{figure}
\pagebreak[4]

\newpage
\begin{figure}[h]
\includegraphics[trim={5cm 2cm 2cm 2cm},scale=0.35]{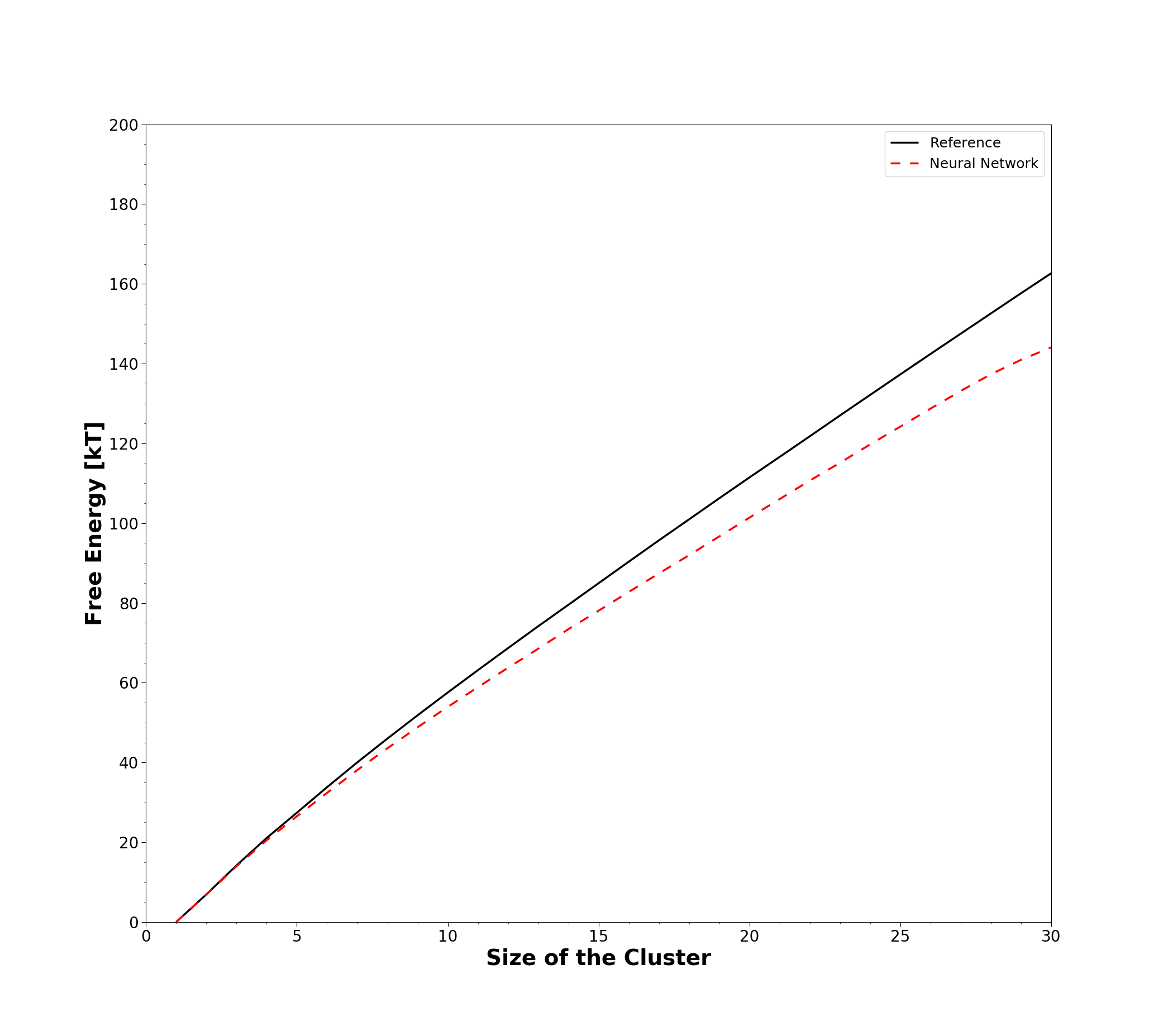}
\caption{\label{ClusterFreeEnergy} The free energy as a function of the number of water molecule's in the cluster. The final neural network's prediction (dashed red line) is plotted on top of the original model's free energy (black line). }
\end{figure}
\pagebreak[4]

\newpage
\begin{figure}[h]
\includegraphics[trim={5cm 5cm 2cm 2cm}, scale=0.35]{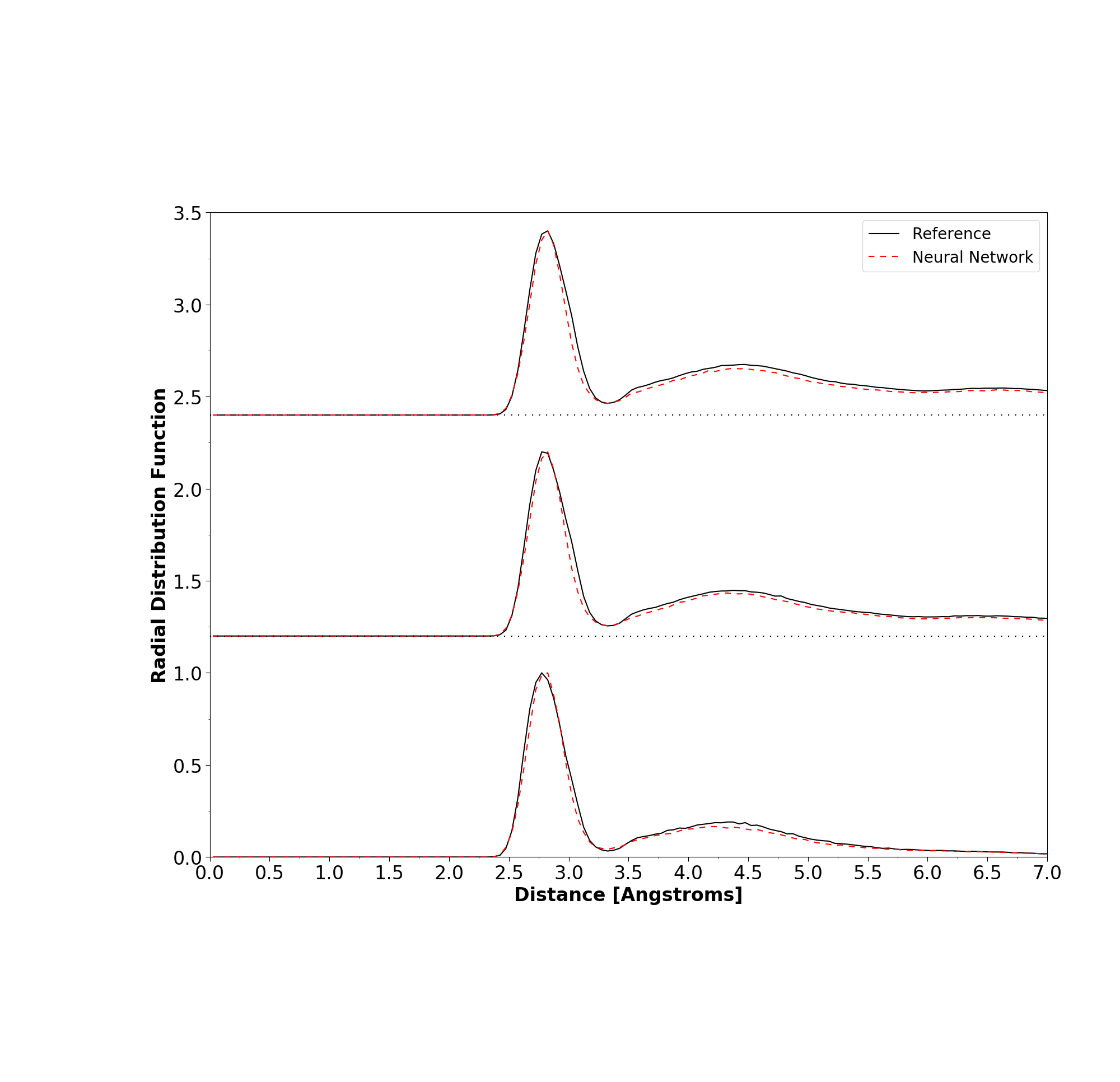}
\caption{\label{ClusterRDF} ANN performance for structural properties. The radial distribution function for a cluster containing 10 water (bottom plot), 30 water (middle plot), and 50 water (top plot) molecules is given here. The final neural network's RDF prediction (dashed red line) is plotted against the original model's RDF (black line). Due to the discrete nature of the system the density tails off to zero and has not been normalized by the system density.}
\end{figure}
\pagebreak[4]

\end{document}


\maketitle

\section{Randomly Generated Training Data}

\begin{figure}[h]
\includegraphics[trim={4cm 0cm 0cm 0cm},scale=0.55]{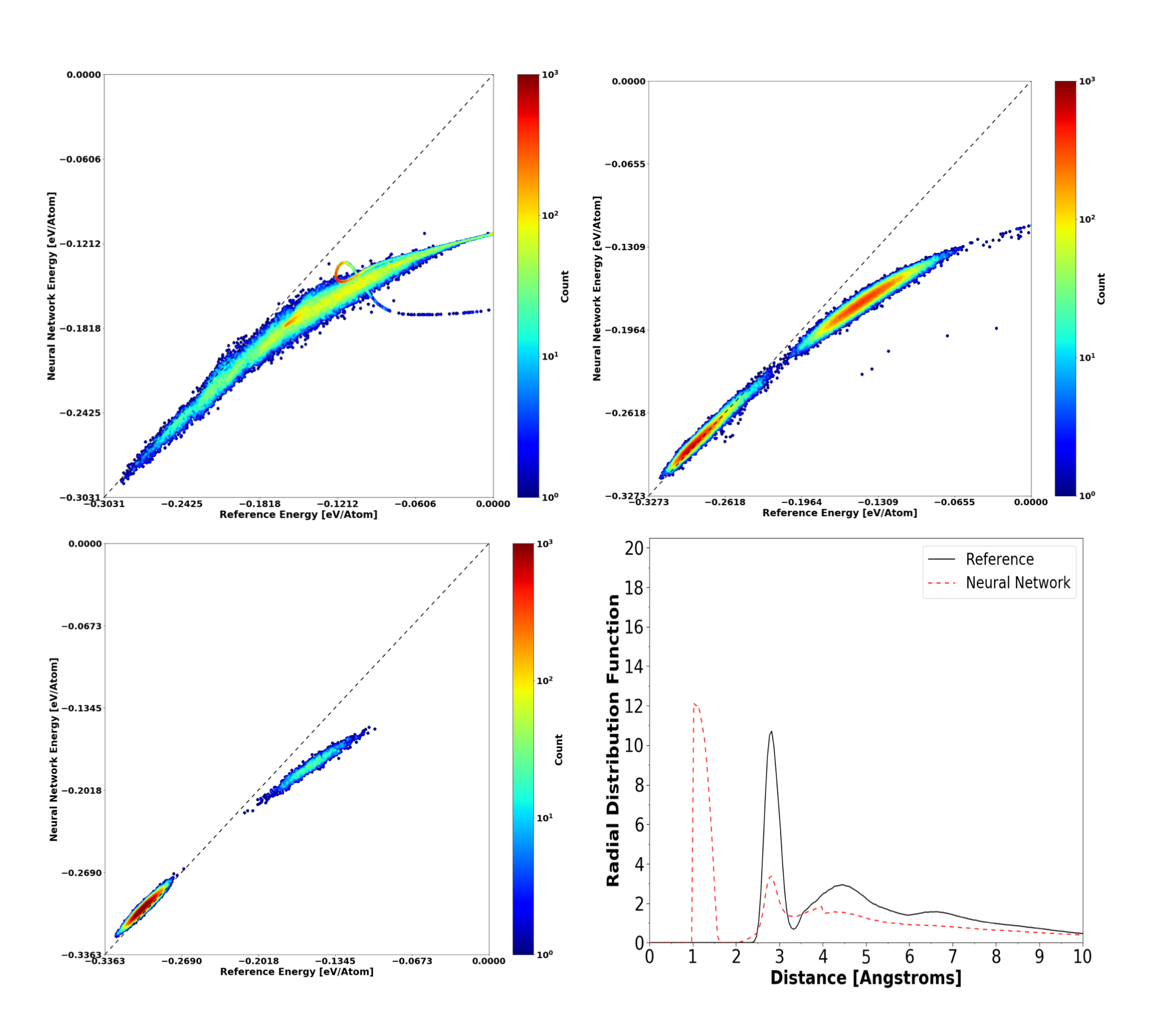}
\caption{\label{RandomNetwork} The results from a neural network trained on a randomly generated set of 426 structures is compared against the same 140,000 structure test set as the AL-ANN potential. The prediction vs reference energies are plotted for the 1-20, 21-50, and 51+ cluster size ranges in the top-left, top-right, and bottom-left panels respectively.  The RDF for a simulation of a fixed cluster size of 50 (dashed red curve) is plotted on top of the reference RDF (solid black line) in the bottom-right panel.}
\end{figure}

A second network with the same network structure was trained on a set of 426 configurations generated via Metropolis MC and Nested Ensemble sampling. This was done to match the final number of structures that the AL-ANN was trained on in order to show what a network trained without the AL scheme would perform.  The results for this network can be found in Fig. \ref{RandomNetwork}.  While according the energy plot the network performs reasonably well for the 20-50 and 51+ range, it is unable to predict the RDF of a cluster size of 50 with any level of accuracy.  It is not sufficient to simply generate the same number of configurations. The configurations must be representative of a wide variety of unique configurations in order to give the network enough information to correctly sample .  

\newpage
\pagebreak[4]